# Assisting Tibetan Students in Learning Quantum Mechanics via Mathematica

Guangtian Zhu[1,2], Jing Hu[3], Chun Du[3*]

[1]Teachers College, Jimei University, Xiamen, China, 361021
[2]College of Teacher Education, East China Normal University, Shanghai, China, 200062
[3]Department of Physics, Tibet University, Lhasa, China, 850000

**Abstract**

*Undergraduate students of physics in Tibet have great difficulty learning quantum mechanics (QM). We attempt to use PER-based methods to help Tibetan students learn QM. In this preliminary study, we incorporate Mathematica in a QM course at Tibet University and record students' learning experiences. Tibetan students tend to have subjective feelings of learning Mathematica, whereas Han students (majority) are more focused on the operational techniques of Mathematica. The results also suggest that both Tibetan students and Han students show limited improvement in time-independent Schrodinger equations after learning QM with Mathematica. Further effort is needed to improve the academic literacy skills of physics students in Tibet.*

## I. Introduction

Tibet is an autonomous minority region in China and has a unique culture. There are 1.3 billion people in China, and approximately 90 percent of Chinese people are from the Han ethnic group. According to the nationwide population census in 2010, there are 6.3 million Tibetan people in China [1]. In Tibet, approximately 2.7 million people are Tibetan, 0.2 million people are Han, and 40 thousand people are other ethnic minorities. The literacy and education level of Tibet's population is lower than the average level in China. Until 2010, approximately 59% of Tibet's population had finished primary school, and only 5.5% people held a university degree in Tibet. Hence, improving the level of education in Tibet is of great concern for educators and governors in China.

Previous research on education in Tibet mostly focused on educational policies, bilingual teaching, or compulsory education for ethnic minority students [2–8]. Few studies have performed physics education research (PER) focused on students in Tibet. One interesting study recorded Tibetan Buddhist monks' learning experiences regarding Electromagnetism. Several members of the PER community were invited to teach physics to a group of overseas Tibetan Buddhist monks [9]. They found that the monks held a unique interpretation about light and colors because "observation of a color is a phenomenon at the boundary between the world and the mind". Such results remind us of the existence of cultural differences in physics education.

To elevate the quality of physics education in Tibet, we endeavor to find a way to improve Tibetan students' understanding of advanced physics courses, such as quantum mechanics (QM). We contacted many students in the physics department at Tibet University about their general difficulties in learning QM. Most of them mentioned three types of difficulties. First, students felt that they could not build connections between an equation and a physical model in QM. When solving problems in QM, students often failed to find an appropriate equation to suit the physical model involved in the problems. Second, they found that the concepts in QM were too abstract to imagine the physical process. Finally, students claimed that they had insufficient foundations in calculus and linear algebra. This weakness in mathematics impeded them from solving differential equations in QM, such as Schrödinger equation. In the interviews, Tibetan physics students expressed their desire for PER tools to reduce their learning difficulties in QM.

Introducing computational tools into physics courses can potentially help students in both conceptual understanding and skills of problem solving [10–13]. As suggested by Chabay and Sherwood, computational activities can improve a physics course in four aspects [13]. First, the computational activities help students connect the formalism of integration in calculus with the procedure of adding up discrete physical quantities. Second, computational tools can better visualize 2-D representations of 3D situations in real world. Third, when modeling physical situations, students can begin with a simplified model in computational tools and then reach a more realistic model by adding features. Last but not least, computational activities provide an opportunity for the students to rethink the merits and underlying goals of physics curriculum.

The PER field have recognized the importance of computational tools for a long time. Through 2005 to 2007, with the sponsorship of AAPT (the American Association of Physics Teachers) and comPADRE (the Communities for Physics and Astronomy Digital Resources in Education), Chonacky and Winch conducted a series of activities for the vision and guidelines for the integration of computation into undergraduate physics. They proposed that "all physics students should be able to write and validate simple programs and be capable of learning on their own to use at least the rudiments of scientific software tools" [14]. In 2018, 1246 faculty from 357 unique institutions in the US responded to a survey about the prevalence and nature of computational instruction in physics courses [15]. The results suggest that more physics faculty have certain computational teaching experience than ten years ago. Out of the 357 departments, 263 departments have at least one faculty reporting experience with teaching computation to undergraduate physics students. Also, 184 departments have at least one faculty member involving computation in advance physics courses such as Quantum Mechanics or Electrodynamics [15].

Various approaches of introducing computational tools into advanced physics courses have been proposed in recent years [16–18]. We were inspired by Dr. Daniel Schroeder's work presented in the AAPT workshop [19]. Dr. Schroeder uses Mathematica software to assist with teaching quantum mechanics at Weber State

University. Mathematica is a computational software developed by Wolfram Alpha that is widely used in fields of math and science [20–25]. For the QM course, Mathematica is a convenient computation tool that relieves students from solving the partial differential equations. Mathematica can also visualize the shape of wave functions so that students can build an intuitive impression of wave functions.

Although there have been many studies regarding the application of Mathematica in physics education, few of them have involved minority students in non-English speaking countries. Because Tibetan students' general difficulties with QM were consistent with the difficulties encountered by students in the US [26–34], we wondered whether the computational activities in advanced physics would be effective for Tibetan students. Therefore, we performed a preliminary study to introduce Mathematica in the QM course for Tibetan students.

In our study, we guided Tibetan students to use Mathematica when solving problems of a one-dimensional infinite square well (1-D infinite square well) in quantum mechanics. Our major research questions include the following points.

- Do the computational activities in the QM course effectively assist Tibetan students in learning 1-D infinite square well? Specifically, how is Tibetan students' performance in a conceptual quiz of 1-D infinite square well compared to the Han students in Tibet University and the American students in public research universities?
- If the Tibetan students perform better or worse than their Han peers in the Mathematical-assisted QM course, what are the possible causes of the difference?

## II. Curriculum Design

We introduced Mathematica to 56 undergraduate students enrolled in a QM course at Tibet University. All the students were majoring in physics, and most of them planned to become secondary school physics teachers in Tibet. Among the 56 students, 30 of them were from the Tibetan ethic group (minority) and 26 of them were Han Chinese (majority). All of these students completed the same QM course taught by the same teacher. Note that the Tibetan students studied QM in the Chinese language because there is no QM textbook written or translated into the Tibetan language. Tibetan students have become accustomed to learning physics in the Chinese language since secondary school.

Few students had previous experience with Mathematica. Hence, we used four additional lessons to teach students the fundamental operations of Mathematica. These four lessons were taught by the same instructor of the QM course in the recitation sessions along with the QM course. Each Mathematica lesson lasted for two hours, and the contents taught in lesson 2 are listed below in italics. The list of contents of all four lessons are provided in the supplementary file (File 2).

- *Major contents covered in the second lesson of Mathematica*
1. *Find the limit of a function*
2. *Calculate an ordinary derivative and a partial derivative*
3. *Calculate an indefinite integral and a definite integral*
4. *Calculate a numerical integral*
5. *Calculate the progression and series*

We adopted the apprenticeship method in teaching students to use Mathematica [35,36]. The instructor first demonstrated how to write Mathematica code to solve problems. Then, tutorials containing sample questions and the corresponding codes were distributed to the students in class. A chapter of the tutorial translated into English is provided in the supplementary files (Files 3-5). Students needed to code in Mathematica to solve calculus problems with reference to the sample codes in the tutorial. The whole class was divided into small groups, each of which contained 2 or 3 students. The instructor and a teaching assistant patrolled the classroom to answer students' questions.

When teaching 1-D infinite square well in the QM class, the instructor still introduced the mathematical procedure of solving a time-independent Schrödinger equation (TISE). The instructor emphasized the physical meaning of normalization and the boundary condition of a wave function in class. Nevertheless, the students were encouraged to solve the TISE with reference to sample codes in Mathematica. By using Mathematica, students can obtain an analytical solution of the TISE as well as the visualized shape of the wave function.

In contrary, in the traditional QM courses taught by the same instructor in the previous semester, students need to learn how to solve TISE in a paper-pencil way. The instructor observed that most students just tried to recite the sinusoidal format of wavefunction for the 1-D infinite square well. When the constant of the square well changed (e.g., width or center of the well), students considered it as a completely new question and failed in solving TISE. Because the burden of quantitatively solving the differential equation was reduced by using Mathematica, we expected the students to have a better qualitative understanding of the QM concepts.

### III. Quantitative Results of the Quiz of 1-D Infinite Square Well

There are several PER-based conceptual tests for upper-level undergraduate QM course, such as Quantum Mechanics Concept Assessment (QMCA), Quantum Mechanics Survey (QMS), Quantum Mechanics Assessment Tool (QMAT), Quantum Mechanics Formalism and Postulates Survey (QMFPS), etc [37–40]. We choose QMS as the question bank because it contained more mathematical expression of calculus. Also, we have QMS data from some universities in the US, which can be used as a reference point for analyzing the Tibetan students' understanding of QM.

We select 6 questions from the Quantum Mechanics Survey (QMS) as a post-test to examine students' conceptual understanding of the time-independent Schrödinger equation (TISE) in a 1-D infinite square well after learning QM with Mathematica [38]. All of the questions are multiple-choice questions without a heavy requirement for calculation. The TISE quiz is provided in the supplementary file (File 6). From the data previously collected in undergraduate QM courses in the US [38], the Cronbach's alpha for the 6-question quiz is 0.75, which suggests that the quiz is reliable. The quiz questions are reviewed by three physics faculty who are familiar with undergraduate quantum mechanics. They confirm that the quiz questions are suitable for the undergraduate QM courses after the students have learned the contents related to TISE in a 1-D infinite square well.

The experimental groups include the 56 students who used Mathematica to assist their learning of QM in 2019. When processing the data, we treated the 26 Han students as experimental group 1 and the 30 Tibetan students as experimental group 2. The control group includes 24 Han students in the 2018 QM class who did not use Mathematica. The control group also took the same conceptual quiz in their semester after they had learned the contents of the 1-D infinite square well. Note that the 2018 Tibetan students are not included in the control group because they studied QM in a different class taught by a different instructor. The 2018 Tibetan students did not take the quiz when learning the content of TISE in 1-D infinite square well.

Table 1. Descriptive statistical data of different groups taking the quiz on the TISE

|  | Sample Size | Mean | Std. Dev. |
|---|---|---|---|
| Control Group (Han 2018) | 24 | 0.92 | 0.909 |
| Exp. Group 1 (Han 2019) | 26 | 1.54 | 1.029 |
| Exp. Group 2 (Tibetan 2019) | 30 | 0.83 | 0.747 |

The descriptive statistical data of the TISE quiz are listed in Table 1. The full score of the six-question quiz is 6. In comparison to the students in Tibet, we calculated the average score of the same six questions from previous QMS data collected in several public research universities in the US (ranking from top 40 to 80 according to the US News). The average is about 1.88 for the students in traditional QM course and 4.64 for the students using PER-based tutorials [38]. None of the three groups in the Tibet University achieved a satisfactory average score on the quiz. The average score of the Han students and Tibetan students in the experimental group is only 1.54 and 0.83 respectively, which is still lower than the average score of the American students in traditional QM course without research-based learning tutorials. However, by analyzing students' answers, we found that students involved in our study did not answer the questions randomly. Their choices fell in several distractors that reflected students' common misconceptions. This result confirms that students in Tibet, whether from the Tibetan or Han ethnic group, are under-represented in physics education. Their academic backgrounds in math and physics are weak, and extra effort is needed to guide them through advanced physics.

Although the absolute score of all three groups was relatively low, we can still find advantage of using Mathematica to assist students' study of QM. The Han students in 2019 who participated in the Mathematica-assisted quantum mechanics course performed significantly better ($p=0.029<0.05$) in the conceptual quiz than the Han students in 2018. The effect size is 0.6. Under the assumption that the Han students in 2018 and 2019 had similar academic backgrounds in math and physics, we can conclude that Mathematica had a positive effect on Han students' learning of quantum mechanics.

Since we do not have the TISE quiz score from the 2018 Tibetan students in the traditional QM class, it is a little bit difficult to direct reflect the improvement of the 2019 Tibetan students who participate in the Mathematica-assisted QM course. However, by comparing the quiz score between 2019 Tibetan students and 2018 Han students, we can infer the positive effect on the Tibetan students when learning QM with Mathematica.

When the Tibetan students and Han students are enrolled in same math/science class taught by the same instructor, the Han students usually performed much better than their Tibetan peers. For the Tibetan and Han students involved in our research, they studied calculus and linear algebra together in their freshmen and sophomore years and the average scores are shown in Table 2. The data in Table 2 suggest that (1) the 2018 Tibetan and 2019 Tibetan students have similar academic competence and (2) there is significant difference ($p<0.05$) between the Tibetan and Han students' performance in calculus and linear algebra. However, the average TISE quiz score of the Tibetan students in 2019 is only slightly lower than their Han peers in 2018 and there is no significant difference ($p=0.723>0.05$) between these two groups. The effect size is only 0.1, which indicates small difference between the 2019 Tibetan students and 2018 Han students. The nonsignificant results between the Tibetan students in the Mathematica-assisted QM course and the Han students in the traditional QM course indicate that the Tibetan students have improved after learning QM with Mathematica.

**Table 2**. Mean and standard deviation of Tibetan and Han students' score (in percentage) of different test. The full score is 100%. For the merit of consistency among different tests, students' scores of the TISE quiz (as in Table 1) are also converted into percentage.

|  | Tibetan 2018 | Han 2018 | Tibetan 2019 | Han 2019 |
|---|---|---|---|---|
| **Calculus** Freshmen Year | 68.9±16.8 | 86.3±9.4 | 70.7±15.1 | 84.9±10.3 |
| **Linear Algebra** Sophomore Year | 67.3±14.9 | 84.9±10.7 | 69.6±16.6 | 84.3±10.9 |
| **TISE Quiz** Junior Year | N/A | 15.3±15.2 | 13.8±12.5 | 25.7±17.2 |

## IV. Qualitative Investigation of Tibetan Students' Uniqueness in the Mathematica-Assisted QM course

Before conducting the preliminary study of applying Mathematica in the QM course, we interviewed many Tibetan and Han students in the Tibet University about their learning difficulties. The claimed difficulties were similar to those expressed by US students, such as connecting the quantum concepts with classical physics, sketching the graph of wavefunctions, differentiating between stationary states and eigenstates, etc [26–31]. Compared to their Han peers, the Tibetan students have much weaker background in math. According to the instructors in the Tibet University, the Han students know the fundamental methods in calculus and linear algebra, such as calculating the cross product of two-by-two matrices or integrating a polynomial, but some of them may not be able to apply these skills correctly. However, many Tibetan students in the QM course are not aware of these fundamental methods, even though they have learned calculus and linear algebra in their freshmen and sophomore years.

We have expected our Mathematica-assisted QM course can relieve Tibetan and Han students from the hassles in math and focus on the conceptual understanding of QM. However, the quantitative result from the 6-question TISE quiz indicates that students' difficulties with QM are not sufficiently compensated in this process, especially for the Tibetan students. In order to probe how to further improve Tibetan students' performance, we conducted qualitative investigation through 1-on-1 interviews and after-class feedbacks. In the qualitative study, Tibetan students exhibit some unique features which may affect their learning experience of the Mathematica-assisted QM course. Our major findings are elaborated as below.

### 1. Tibetan students have lower proficiency with coding in Mathematica compared to their Han peers

At the end of the semester, we conducted 1-on-1 interviews with Tibetan students and Han students about their experiences learning QM with Mathematica. The interview was semi-structured in a think-aloud protocol. We showed a QM question of 1-D infinite square well and a calculus question to the interviewees. A laptop is also provided to the students and they are asked to use Mathematica when solving the problems. After the students finished solving the QM question and the calculus question, we further interviewed them with some questions of learning experience, such as their impressions about Mathematica, their pains and gains in studying QM with Mathematica, and their advices on Mathematica-assisted QM courses.

Six Tibetan students and six Han students participated the 1-on-1 interview. None of these students had used Mathematica before. The Tibetan and Han students expressed similar interest in Mathematica in the interviews but their proficiencies of coding with Mathematica were different. In the interview, students need to solve an ordinal differential equation in Mathematica. Such content had been practiced several times in the Mathematica-assisted QM course. Four Han students were able to write the proper codes in Mathematica whereas only two Tibetan students successfully did so. For the four Tibetan and two Han students who cannot write the codes, we showed them sample codes and asked them to explain. The two Han students can correctly explain the sample codes. However, only one Tibetan student interpreted the content of the sample codes. The other three Tibetan students could not recognize the meaning of some symbols in the sample codes.

A Tibetan student claimed that his proficiency with computers was not as good as his Han peers because he has no computer at home. However, it is noteworthy that the lack of a computer at home was not due to poverty, at least for the student involved in the interview. Approximately 60% of university students in Tibet are children of farmers or nomads. Tibet University provides sufficient need-based scholarships for these undergraduate students to cover their tuition and living costs. The student without a personal computer in the Mathematica class is from a nomad family that moves from place to place in different seasons. Hence, a computer is not considered a necessity in his family. This student typically used the library computer at the university to finish his homework.

## 2. Tibetan students report more feedback of personal feelings than feedback of operational techniques

Besides the 1-on-1 interview, we also collected students' feedback after each Mathematica-assisted QM lesson. All students in the course (not restricted to the 12 students involved in the 1-on-1 interview) were encouraged to submit feedback about their personal learning experience with Mathematica. In general, students' experience can be divided into two categories. One category is about operational techniques while using Mathematica, and the other focuses on students' emotional feelings while working on Mathematica assignments. For example, we have collected 32 pieces of effective feedbacks after the first two Mathematica lessons. Fifteen pieces of these feedbacks are emotional feelings and seventeen pieces are operational techniques. Below we present part of the feedbacks that Tibetan students and Han students submitted. The full list of feedbacks is provided in the supplementary file (File 7).

- Some feedback on operational techniques (mostly provided by Han students)

*…A space should be added between y'[x] and x.*
*…The code "TraditionalForm" can simplify the output.*
*…The expressions "==" and "=" have different meaning.*
*…If Mathematica does not deliver an output correctly, saving the file and restarting the computer may work.*
*…The correct way of expressing sin3x is (Sin[x])^3. The wrong way is (Sin^3)[x].*
*…The arrows must be selected from the Basic Math Assistant Palette.*
*…In order to input a fraction, we need to type dy first and then press the button for a*

*fraction.*

- Feedback on emotional feelings (mostly provided by Tibetan students)

*…We do not understand the contents in the tutorial.*
*…We cannot build connections between the sample questions and the exercises.*
*…We did not get the graph in the first trial. Then, we wrote the code again, and our hard work paid off.*
*…The second class was easier than the first one. We did not have many problems except that we were slow in typing the symbols. We discussed together to successfully finish the assignment.*
*…The course was difficult, but we successfully finished the assignment by discussing with each other.*
*…We had difficulty in the assignment because the sample question in the tutorial was not the same as the assignment.*
*…We were not familiar with the software, and sometimes we had errors in missing symbols. After a careful examination, we corrected the errors.*

Compared to the Han students, Tibetan students reported more emotional experiences than operational experiences when learning Mathematica. Among the 15 pieces of feedbacks on emotional feelings, 11 pieces were provided by Tibetan students. Also, only 6 out of 17 pieces of feedbacks on operational techniques were provided by Tibetan students. Tibetan students' expressions of operational experience were often vague, which may impede their improvement in learning Mathematica. For example, Tibetan students mentioned that they "encountered problems in details during the operation process". However, they did not specify what kind of problem they met or how they solved the problem. By contrast, as shown above, many Han students precisely recorded the hints they used to solve their problems with operating Mathematica.

The difference between Tibetan and Han students in expression of feedback is also observed by other professors teaching advanced physics curriculum. For example, in the courses of Classical Mechanics or Electrodynamics, when the professors ask students why they cannot solve a problem, the typical feedback from Tibetan students is usually "too hard" or "do not understand". Yet the Han students may tell the professors they get stuck in a particular step of solving an equation or they cannot integrate certain function. Such results remind us that providing educational tools or techniques alone is far from sufficient to enhance minority students' learning processes. The instructors need to make an effort in training students' academic literacy, e.g., describing problems concretely for others to understand or making notes of learning experiences for future reference. Grasping such academic literacy is as important as mastering the physics content for minority students.

3. **Tibetan students' learning attitudes are influenced by social-cultural factors**

For historical and geographic reasons, the average number of university students per thousand residences in Tibet is only 55/1000, which is much lower than that of mainland China, i.e., 89/1000. Therefore, the government and universities in Tibet have a strong urge to increase the number of university students. The number of graduates is a key performance index for administrators. In such circumstances, it would be a hassle for instructors and department chairs to fail a student on a final exam, which may increase the risk of a decreasing graduation rate. Hence, many instructors are willing to narrow the content range of a final exam, create easy questions that are almost exactly the same as examples in textbooks, and assign a large amount of partial credits for students' incorrect answers.

Due to the simplicity of final exams and the generous grading rubric, university students in Tibet have generated two types of attitudes about learning physics. On the one hand, some students mistakenly consider physics an easy subject and believe they only need to make an effort at the last moment. In the 1-on-1 interview after the QM course, one student proudly told us he "mastered" all the content regarding thermodynamics the night before the final exam and successfully passed the exam. This student used the example to explain why he cannot solve the QM problem we provided in our interview. He claimed that he could definitely solve the QM problem if he only had one hour to study the contents.

On the other hand, some students admitted that they cannot understand the content taught in class and stated that they gave up on following the instructor's lectures in class. These students knew that they could pass exams in most cases as long as they were not the worst in the class. Even if they failed an exam, they still had a great chance to pass the makeup test. Hence, these students did not have sufficient motivation to learn the difficult contents in advanced physics.

Introducing Mathematica in the QM class partially increases students' motivation to study. Several Tibetan students claimed that they at least learned some contents in the Mathematica class because they needed to participate in the group study and operate the software. Moreover, because most of the students in the physics department at Tibet University plan to be physics teachers in Tibet, they are aware of their duty as teachers. A Tibetan student mentioned that he was willing to learn Mathematica so that he could be more competitive than the senior teachers at his future school. Another Tibetan student also expressed her plan to use Mathematica to draw graphs in the high school physics courses she is going to teach.

Many Tibetan students claimed that they successfully finished assignments by collaboration or discussion with their teammates. The preference for teamwork was also revealed in previous studies of Tibetan students' thinking styles [3]. There may be some socioeconomic reasons for Tibetan students' tendency for teamwork. As mentioned before, many Tibetan students are children of farmers or nomads, and it is necessary for their families to help each other in their neighborhood to survive in the rigorous environment. Tibetan's religious belief of Buddhism also treasures the value of helping other people. Therefore, collaborative study should be incorporated in Tibetan students' learning process.

V.  **Summary**

In this preliminary study, we used Mathematica to assist in a QM course for students in Tibet. The performance of Tibetan and Han students improved on a test of a 1-D infinite square well. However, the overall effect of Tibetan students' learning QM with Mathematica was not very satisfactory. Tibetan students' weakness in learning advanced physics was due to multiple factors, e.g., insufficient mathematical background, lack of proficiency in computer operation, and low pressure of passing the exam.

In the manuscript, we also record Tibetan students' feedbacks when they were involved in the Mathematica-based QM course. When applying the new educational tools to minority students, instructors may encounter situations different from practice in majority groups. Hence, we believe that elaboration of Tibetan students' reactions in the learning process will be helpful for the minorities in many regions.

As indicated by Tibetan students' feedback about their learning experience, Tibetan students had deficiencies in clearly recording the difficulties they encountered and how those difficulties were solved. In follow-up courses, we plan to guide Tibetan students to properly recording their learning experiences with Mathematica. We will study whether proper guidance can enhance Tibetan students' learning habits and improve their retention of Mathematica operations.

**Acknowledgements**

This research compiles with the IOP Science ethical standards in the treatment of subjects, and has been approved by the University Committee on Human Research Protection in the East China Normal University (approval number HR 055-2018). We have also obtained consent from all identifiable individuals involved in the research. This project is supported by the Research Funds of Happiness Flower ECNU (2020ECNU-XFZH007) and the Tibet University. Any opinions, findings, and conclusions expressed in this paper are those of authors and do not necessarily reflect the views of the funding agencies.